\documentclass[preprint]{aastex}
\usepackage{graphicx}

\usepackage{natbib}
\usepackage{wrapfig}

\usepackage{graphics}

\usepackage{natbib}

\newcommand{\be}{\begin{equation}}
\newcommand{\ee}{\end{equation}}
\newcommand{\nn}{\mbox{} \nonumber \\ \mbox{} }
\newcommand{\ba}{\begin{eqnarray}}
\newcommand{\ea}{\end{eqnarray}}

\newcommand\eg{\textit{e.g.,\ }}

\def\aap{A\&A}                

\begin{document}

\title{On the sideways expansion of  relativistic non-spherical shocks and  GRB afterglows}




\author{Maxim Lyutikov\\
Department of Physics, Purdue University, \\
 525 Northwestern Avenue,
West Lafayette, IN
47907-2036 }

\begin{abstract}
Expansion of non-spherical relativistic blast waves is considered in the
Kompaneets (the thin shell) approximation.  We find that the relativistic motion effectively
``freezes out'' the lateral dynamics of the shock front: 
only extremely strongly collimated shocks, with the opening 
angles $\Delta \theta \leq 1/\Gamma^2$, 
show  appreciable modification of profiles due to sideways expansion.
For less collimated profiles the propagation is nearly ballistic;    the sideways expansion of  relativistic shock 
becomes important only when they become mildly  relativistic.
\end{abstract}

\maketitle

\section{Introduction}

Dynamics and corresponding radiative signatures of  non-spherical relativistic shocks remains an important unresolved issues in studies on Gamma Ray Bursts (GRBs).
Since GRBs produce narrowly collimated outflows that evolve laterally, understanding the overall dynamics - both theoretical and in terms of  agreement between different numerical results - is imperative to the interpretation of the broadband observations of GRBs \cite{Rhoads99,Frail01}.

Presently, there are two  competing views on the lateral evolution  of the relativistic outflows. Theoretically, it is typically  
argued that the lateral evolution of the flow proceeds with relativistic velocities  \citep{PiranReview}, \citep[see also][]{2011arXiv1102.5618W}. This view is contradicted by the results of numerical simulations that show very little lateral evolution in the relativistic regime \cite{2004ApJ...601..380C,2009ApJ...698.1261Z,2010A&A...520L...3M,2011arXiv1105.2485V}.

In this Letter we argue that this disagreement results from the incorrect theoretical assumptions about the lateral evolution of the flow.  
What is important for the interpretation of observations  is the  evolution of a curved shock.  Previously    
  the  lateral evolution of the non-spherical shocks was   incorrectly treated   as a free lateral  expansion into vacuum \citep[\eg][Eq. 5]{2011arXiv1102.5618W}..   The assumption of
the   lateral expansion  with the  sound speed  results in a  ``gramophone-type'' profiles and
{\it  exponential }  slowing down of the ejecta. This has drastic implications for the underlying light curves \citep[eg][]{2000ApJ...541L...9K,2003ApJ...592..390P}. In fact the dynamics of the non-spherical shocks is more subtle; the correct treatment, as we argue below, is consistent with slow lateral evolution seen in numerical simulations.

Evolution  of strong non-spherical shocks is a  well studies  problems in fluid dynamics.
The two fundamental works that have laid the foundation 
 for non-spherical (two-dimensional) shocks, due to   \cite{Komp}  and to 
 \cite{LaumbachProbstein},  were originally designed  to treat strong shock
waves in the non-isotropic medium.
These two complimentary methods
 have been extensively applied in  astrophysics
to treat supernova explosions \citep{Bisnovatyi-KoganSilich95} and non-isotropic winds \citep[\eg][]{Icke}.
In the  Kompaneets approximation the internal pressure of the
gas is assumed to be constant. Then the Rankin-Hugonio conditions
determine the normal velocity of the shock in the external inhomogeneous
medium.  A modification of 
the  Kompaneets approximation - a thin or snowplow shell approximation -
 has also been used extensively  \cite[\eg][]{1978ApJ...221...41W,1988ApJ...324..776M,1989Ap&SS.154..229B}. In a complimentary    Laumbach-Probstein approach  \citep {LaumbachProbstein} the streamlines of the shocked material
are assumed to be radial, thus   neglecting the 
lateral pressure forces.

The relativistic generalization of the
  Kompaneets  and the Laumbach-Probstein methods have been discussed by  \cite{1979ApJ...233..831S}. 
     Relativistic  dynamics provide extra support
for the thin shell method, since in the relativistic blast waves
the shocked material is concentrated in even narrower region
$R/\Gamma^2$ than in the non-relativistic Sedov solution.
In addition,  the limited causal connection (over the angle $\sim 1/\Gamma$)
provides a justification for the Laumbach-Probstein method on the
angle scale comparable to $1/\Gamma$.
As has been pointed out by  \cite{1979ApJ...233..831S}, the two methods - Kompaneets and
Laumbach-Probstein - become  very similar in the relativistic regime.
This is due to the fact the in a relativistic quasi-spherical wave, 
the typical angle that a shock wave makes with the direction
of the velocity is of the order $ \alpha \sim 1/\Gamma^2$.
Thus the post shock pressure along the shock differs only by one part
in $\Gamma^2$, so that both approximations of
 constant post-shock pressure
and radial post-shock motion become equivalent.
 In some sense  the propagation of 
 strongly  relativistic non-spherical shocks becomes trivial:
relativistic kinematic effect freeze out the lateral dynamics of the flow so 
that
different parts of the flow behave virtually independently.

\section{Relativistic non-spherical  shocks in the  thin shell  approximation}

In this section we re-derive the relativistic Kompaneets equation  \cite{Komp,1979ApJ...233..831S} allowing for the arbitrary
velocity of the shock and arbitrary (angle-dependent) luminosity
and/or external density. 
Consider a shock propagating with  a three-velocity $V$ at an angle
$\alpha$ to its normal. There are three generic rest frames
in the problem: laboratory frame $K$, a frame where the shock is normal to the
flow $K_1$ and a shock rest frame $K_0$. 
A frame $K_1$ is related to the lab frame $K$ by a Lorentz boost along 
$y$ axis with a Lorentz factor $\Gamma_\parallel  = 
1/\sqrt{1- V^2 \sin^2 \alpha}$. 
In $K_1$ the velocity of the shock is 
$V_1 = \Gamma_\parallel V \cos \alpha $ (along the $x$ direction). 
Thus, $\Gamma_1^2 = 1/(1-V_1^2) =  \Gamma^2 \cos^2  \alpha + \sin ^2  \alpha$
(the shock becomes non-relativistic when $\pi/2-  \alpha \sim 1/\Gamma$).
In the frame $K_0$ $V_1$ and $\Gamma_1$ are the velocity and the Lorentz
factor of the unshocked medium.
 In the lab frame 
the $x$ component of the shocked velocity
$V_x ' = V_1' /( \Gamma_\parallel (1+  V_1' V \sin \alpha)$
 generally has a completed
form, but simple relations can be obtained in the strongly relativistic
limit (see below).

We introduce next an acceleration parameter $K$ (Kompaneets 1960, Icke 1988)
as a Lorentz factor of the shock in the $K_1$ frame
\be
K = \Gamma_1^2
\ee
The acceleration parameter can be expressed in terms of a ratio
of a post shock pressure $P'$  to the upfront density $\rho$.
For  relativistic  strong shocks  with the ratio of specific heats
$\hat{\gamma}=4/3$
\be
K= {2\over 3} {P' \over \rho} 
\ee
while for non-relativistic shock with  $\hat{\gamma}=5/3$
\be
V_1^2 = 1-{ 1\over K}  = 
{4 \over 3}  {P' \over \rho}
\ee
 
Expressing the relevant quantities in terms of $K$ we find
\ba &&
\Gamma = \sqrt{K -  \sin ^2  \alpha \over \cos^2  \alpha } 
\approx  { \sqrt{K } \over  \cos  \alpha }
\nn &&
V = \sqrt{K-1 \over K -  \sin ^2  \alpha } 
\approx 1-  {  \cos^2  \alpha \over 2  K}
\nn &&
V_1= \sqrt{1-{1\over K}} \approx 1-{1\over 2 K}
\nn &&
V_1' = \sqrt{ 1-{2 \over K}}  \approx 1-{1\over K}
\nn &&
\Gamma_\parallel = \sqrt{K - \sin^2 \alpha \over K \cos ^2 \alpha } 
\approx {1 \over \cos \alpha}
\nn && 
V_x' \approx
{ \cos \alpha \over 1 + \sin \alpha} 
\left( 1 - { (2+ \sin \alpha) \cot^2 \alpha/2 \over 2 K} \right),
\ea
where the approximations assume strongly  relativistic motion.

Finally, 
expressing $V $ in terms of $V_1$ we find
\be
V^2 = { V_1^2 \over  V_1^2 \sin ^2  \alpha + \cos^2 \alpha}  =
{ 1 - {1 / K} \over 1- \sin^2  \alpha /K }
\approx \left\{
\begin{array}{cc}
{V_1^2 \over \cos^2 \alpha}  & \mbox{ if $V_1 \ll 1$} \\
1 - { \cos^2 \alpha \over K}  & \mbox{ if $V_1 \rightarrow 1$, 
arbitrary $\alpha$}\\
  \left(1 - {1 \over K}\right) \left( 1 + { \sin ^2  \alpha \over K} \right) 
 & \mbox{ if $ \alpha \rightarrow 0$}
\end{array} \right.
\ee

\begin{figure}[h!]
\includegraphics[width=\linewidth]{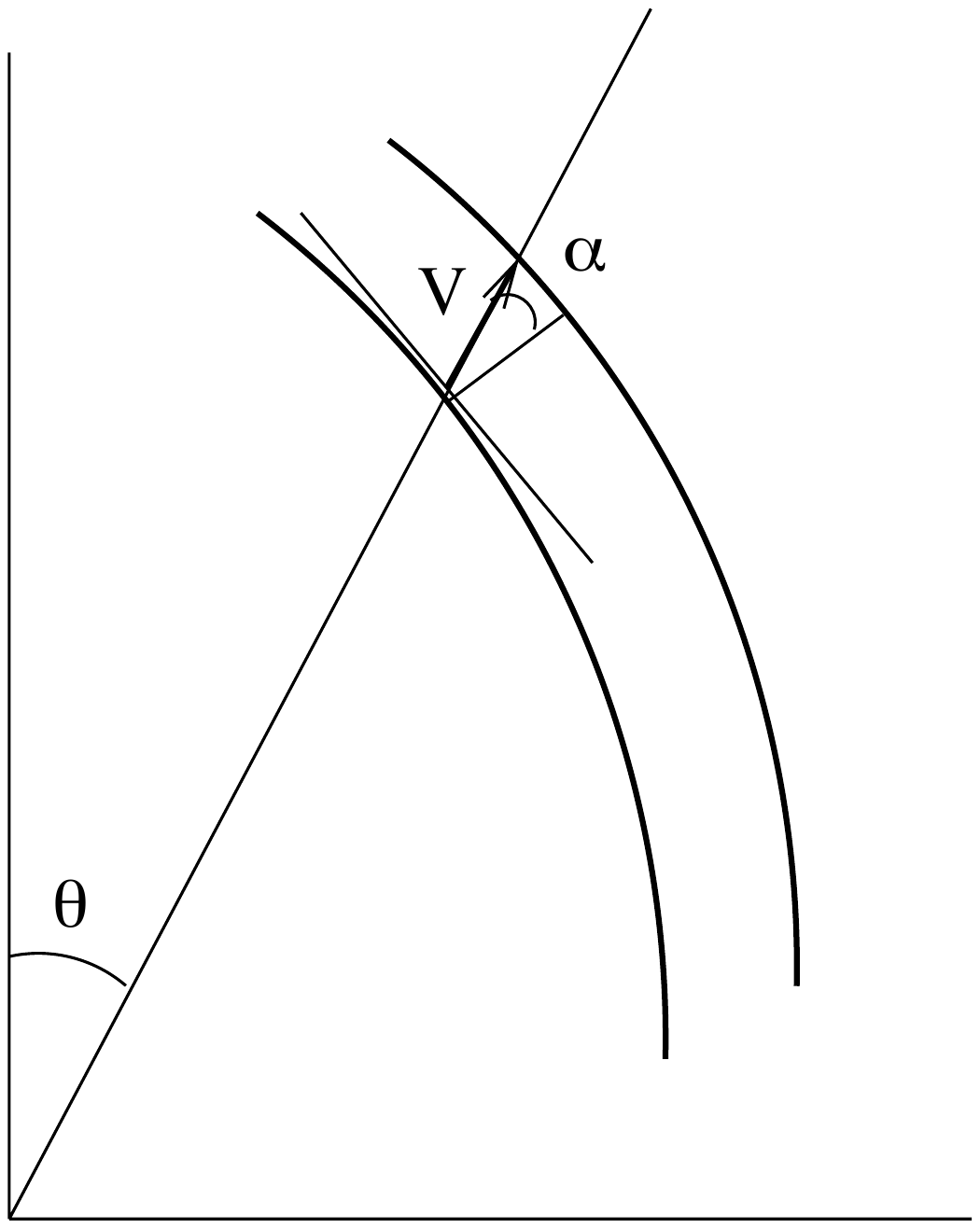}
\caption{Geometry of the  flow. The central source located at the origin produces anisotropic wind with luminosity depending on the polar angle $\theta$.   At a polar angle $\theta$ the shock is located at radius $R(\theta)$, while the wind direction (radial direction) makes an angle $\alpha$ with the shock normal}
\label{obliq1}
\end{figure}

Consider  next a small section  of the non-spherical shock at the spherical polar
angle $\theta$ propagating at an angle 
\be
\tan \alpha = - { \partial \ln R \over \partial \theta}
\ee
to the radius vector (Fig. \ref{obliq1}).
Then
\be
 \left({ \partial R \over \partial t} \right)^2 = V^2 =
{ 1 - {1 / K} \over 
1-  {( \partial_\theta \ln R)^2  \over K (1 + ( \partial_\theta \ln R)^2) } }
\label{Komp}
\ee
Here $K$ is a function  of the shock position and angle
$K\equiv K(\theta, R)$.
Equation (\ref{Komp}) is the sought relativistic generalization
of the Kompaneets equation (the thin shell modification of the
Kompaneets equation, to be more precise).
For example, for non-relativistic $V = 1-1/K \ll 1$
eq. (\ref{Komp}) reduces to  the familiar  Kompaneets form
\be
{ \partial R \over \partial t}  = 
V \sqrt{  (1 + ( \partial_\theta \ln R)^2) } 
\ee

The other two 
simplifying cases of the  Kompaneets equation include
 relativistic motion, $K  \gg 1 $:
\be
 \left({ \partial R \over \partial t} \right)^2 =
 1 - {1 \over  K} 
\left( { 1 \over 1+ ( \partial_\theta \ln R)^2}  \right)
\ee
and arbitrary quasi-spherical  motion, $\alpha \ll 1$, $K$-arbitrary
\be
 \left({ \partial R \over \partial t} \right)^2 =
\left( 1- {1 \over  K} \right)
\left( 1+ {1 \over  K}  ( \partial_\theta \ln R)^2  \right)
\ee
Again,
the last equation readily gives the standard Kompaneets equation
as $K\rightarrow 1$ and $ 1- 1/K \rightarrow V^2$.

Examination of the eq. (\ref{Komp}) confirms that
for relativistic motion the angle $\alpha \sim 1/(\Gamma^2 \Delta \theta)
\sim 1/(K  \Delta \theta) $
(where $\Delta \theta$ is a typical angular scale for a change in a Lorentz
factor. Thus, unless $\Delta \theta \sim 1/ \Gamma^2 \sim 1/ K $, 
the term $ ( \partial_\theta \ln R)^2 \sim 1/K^2$ is of much 
higher order in $1/K$ and can be  neglected.
This express the fact the the lateral dynamics of strongly relativistic 
shock waves (in fact of any strongly relativistic motion)
is ``frozen out'' by kinematic effects.

If the shock is not strongly collimated,  $\Delta \theta \gg  1/\Gamma^2$,
 we can 
neglect the factor $ \partial_\theta \ln R $, different parts of the shock
will propagate radially with a different Lorentz factor given
by the driver or the external density inhomogeneity:
\be
 \left({ \partial R(t, \theta) \over \partial t} \right)^2 = 
{ 1 - {1 / K (t, \theta) } }
\ee
This approximation is similar to  \cite{LaumbachProbstein}   approximation,
which assumes radial motion.
 Thus, in  the strongly relativistic case both
Kompaneets and Laumbach-Probstein
become equivalent, consistent with the conclusion of \cite{1979ApJ...233..831S}.

In the strongly relativistic regime, $K\equiv \Gamma_1^2 \gg 1$,
the lateral dynamics of the flow is frozen out, unless the flow
is extremely strongly collimated with $\Delta \theta \leq 1/ \Gamma_1^2$.
Keeping  $( \partial_\theta \ln R)$ arbitrary and expanding in $1/K$
 eq. (\ref{Komp}) takes the form
\be
 \left({ \partial R \over \partial t} \right)^2 \sim
 1 - {1 \over  K} \left(  {  1 \over 1+
( \partial_\theta \ln R)^2 } \right) 
\label{Komp1}
\ee


\section{Discussion}

In this paper we considered the lateral evolution of non-spherical relativistic outflows. Contrary to the commonly assumed fast  lateral expansion,  we find that unless the shape of the shock is extremely narrow, with the opening angle of the order of $1/\Gamma^2$, the lateral evolution is effectively frozen out by the highly relativistic motion of the shock. 
Thus, we confirm the conclusion of   \cite{1979ApJ...233..831S} that highly relativistic shock propagate nearly ballistically. Our conclusion is  broadly consistent with the results of numerical simulations showing very slow  \citep[logarithmic, \eg][]{} lateral evolution. In contrast, the calculations of the afterglow emitted spectra have to be reconsidered accordingly.

I would like to thank Dale Frail, Hendrik van Eerten, Andrew MacFadyen and Eli Waxman for discussion.
 
\bibliographystyle{apj}

  \bibliography{/Users/maxim/Home/Research/BibTex}

\end{document}